\documentclass[letter,traditabstract,longauth]{aa} 
\usepackage{graphicx}
\usepackage{txfonts}
\usepackage{natbib}
\bibpunct{(}{)}{;}{a}{}{,} 
\begin{document}

   \title{The zCOSMOS redshift survey: \\
   evolution of the light in bulges and discs since $z \sim 0.8$
   \thanks{Based on observations obtained at the European Southern
   Observatory (ESO) Very Large Telescope (VLT), Paranal, Chile, as part of the
   Large Program 175.A-0839 (the zCOSMOS Spectroscopic Redshift Survey)}}

   \author{L. A. M. Tasca\inst{1}
   \and L.~Tresse  \inst{1}
   \and O.~Le F\`evre  \inst{1}
   \and O.~Ilbert \inst{1}
   \and S.~J. Lilly  \inst{2}
   \and G.~Zamorani \inst{3}
   \and C.~L\'opez--Sanjuan \inst{4}
   \and L.~C. Ho \inst{5,6}
   \and S.~Bardelli \inst{3}
   \and A.~Cattaneo \inst{1}
   \and O.~Cucciati \inst{3}
   \and D.~Farrah \inst{7}
   \and A.~Iovino \inst{8}
   \and A.~M. Koekemoer \inst{9}
   \and C.~T.~Liu \inst{10}   
   \and R.~Massey \inst{11}
   \and A.~Renzini \inst{12}
   \and Y.~Taniguchi\inst{13}
   \and N.~Welikala \inst{14}
   \and E.~Zucca \inst{3}
   \and C. M. Carollo \inst{2}
   \and T.~Contini \inst{15,16}
   \and J.-P. Kneib \inst{17}
   \and V.~Mainieri \inst{18}
   \and M.~Scodeggio \inst{19}
   \and M.~Bolzonella \inst{3}
   \and A.~Bongiorno \inst{20}
   \and K.~Caputi \inst{21}
   \and S.~de la Torre \inst{1}
   \and P.~Franzetti \inst{19}
   \and B.~Garilli \inst{19}
   \and L.~Guzzo \inst{8}
   \and P.~Kampczyk  \inst{2}
   \and C.~Knobel \inst{2}
   \and K.~Kova\v{c} \inst{2}
   \and F.~Lamareille \inst{15,16}
   \and J.~-F.~Le Borgne \inst{15,16}
   \and V.~Le Brun \inst{1}
   \and C.~Maier \inst{2}
   \and M.~Mignoli \inst{3}
   \and R.~Pello \inst{15,16}
   \and Y.~Peng \inst{2}
   \and E.~Perez Montero \inst{15,16,22}
   \and R. M. Rich \inst{23}
   \and M.~Tanaka \inst{24}
   \and D.~Vergani \inst{25,3}
   \and R.~Bordoloi \inst{2}
   \and A.~Cappi \inst{3}
   \and A.~Cimatti \inst{26}
   \and G.~Coppa \inst{19}
   \and H.~J. McCracken \inst{27}
   \and M.~Moresco \inst{3}
   \and L.~Pozzetti \inst{3}
   \and D.~Sanders \inst{28}
   \and K.~Sheth \inst{29}
          }

   \offprints{L.A.M. Tasca}

   \institute{Aix Marseille Universit\'e, CNRS, LAM 
   (Laboratoire d'Astrophysique de Marseille) UMR 7326, 13388, 
   Marseille, France \\
              \email{lidia.tasca@oamp.fr}
   \and
   Institute of Astronomy, ETH Zurich, CH-8093, Zurich, Switzerland
   \and
   INAF Osservatorio Astronomico di Bologna, via Ranzani 1, I-40127, Bologna, Italy
   \and
   Centro de Estudios de F\'isica del Cosmos de Arag\'on, Plaza San Juan 1, planta 2, 44001 Teruel, Spain
   \and
   Kavli Institute for Astronomy and Astrophysics, Peking University, Beijing 100871, China
   \and
   The Observatories of the Carnegie Institution for Science, 813 Santa Barbara Street, Pasadena, CA 91101, USA
   \and
   University of Sussex, Falmer, Brighton BN1 9QH, UK
   \and
   INAF Osservatorio Astronomico di Brera, Via Brera 28, I-20121 Milano, Italy
   \and
   Space Telescope Science Institute, 3700 San Martin Drive, Baltimore, MD 21218
   \and
   Astrophysical Observatory, City University of New York, College of Staten Island, 2800 Victory Blvd, Staten Island, NY, 10314, USA
   \and
   Institute for Computational Cosmology, Durham University, South Road, Durham
   \and
   INAF - Osservatorio Astronomico di Padova, Padova, Italy
   \and
   Research Center for Space and Cosmic Evolution, Ehime University, Bunkyo--cho, Matsuyama 790--8577, Japan
   \and
   Department of Physiscs, Oxford University, Denis Wilkinson Building, Keble Road, Oxford OX1 3RH, UK
   \and
   Institut de Recherche en Astrophysique et Planetologie, CNRS, 14, avenue Edouard Belin, F--31400 Toulouse, France
   \and
   IRAP, Universite de Toulouse, UPS--OMP, Toulouse, France
   \and
   LASTRO, Ecole polytechnique f\'ed\'erale de Lausanne, Suisse
   \and
   European Southern Observatory, Karl-Schwarzschild-Strasse 2, Garching, D-85748, Germany
   \and
   INAF--IASF, Via Bassini 15, I-20133, Milano, Italy	 
   \and
   Max--Planck--Institut fur extraterrestrische Physik, Giessenbachstrasse, D--85748 Garching bei Munchen, Germany
   \and
   Kapteyn Astronomical Institute, University of Groningen, P.O. Box 800, 9700 AV Groningen, The Netherlands
   \and
   Instituto de Astrofisica de Andalucia, CSIC, Apartado de correos 3004, 18080 Granada, Spain
   \and
   Department of Physics and Astronomy, University of California, Los Angeles, CA 90095--1547, USA
   \and
   National Astronomical Observatory of Japan, 2-21-1 Osawa, Mitaka, Tokyo 181-8588, JAPAN 
   \and 
   INAF--IASF Bologna, Via P. Gobetti 101, I--40129 Bologna, Italy
   \and 
   Dipartimento di Astronomia, Universit\`a degli Studi di Bologna, Bologna, Italy
   \and
   Institut d'Astrophysique de Paris, UMR 7095 CNRS, Universit\'e Pierre et Marie Curie, 98 bis Boulevard Arago, F--75014 Paris, France
   \and
   Institute for Astronomy, 2680 Woodlawn Drive, Honolulu, HI 96822-1839, USA
   \and
   North America ALMA Science Center, National Radio Astronomy Observatory
   }

   \date{Received 23 February 2014 ; accepted 13 March 2014}

 
  \abstract
   {We studied the chronology of galactic bulge and disc formation by analysing
   the relative contributions of these components to the $B$--band rest--frame
   luminosity density at different epochs.
   We present the first estimate of the evolution of the fraction of rest--frame
   $B$--band light in galactic bulges and discs since redshift $z\sim 0.8$. 
   We performed a bulge--to--disc decomposition of HST/ACS images of 3266 
   galaxies in the zCOSMOS--bright survey with spectroscopic redshifts in the
   range $0.7 \leq z \leq 0.9$.
   We find that the fraction of $B$--band light in bulges and discs is 
   $(26 \pm 4)\%$ and $(74 \pm 4)\%$, respectively.
   When compared with rest--frame $B$--band measurements of galaxies in the
   local Universe in the same mass range 
   ($10^{9} M_{\odot}\lessapprox  M \lessapprox 10^{11.5} M_{\odot}$), we find
   that the $B$--band light in discs decreases by $\sim30\%$ from
   z$\sim 0.7-0.9$ to z$\sim0$, while the light from the bulge increases by
   $\sim30\%$ over the same period of time.   
   We interpret this evolution as the consequence of star formation and mass 
   assembly processes, as well as morphological transformation, which gradually
   shift stars formed at half the 
   age of the Universe from star--forming late--type/irregular galaxies to
   earlier types and ultimately into spheroids.  
   }
   

   \authorrunning{Lidia A. M. Tasca}
   \titlerunning{Evolution of Bulges and Discs since z$\sim 0.8$}

   \maketitle


\section{Introduction}

Several physical processes are at work to assemble mass and shape galaxies 
during cosmic time, but their relative contributions and effective time--scales
are as yet unclear. 
In the hierarchical dark--matter halo assembly picture, galaxies obtain their baryonic mass 
through different processes that include major and minor mergers or continuous 
gas accretion, and lose mass when subject to strong feedback from 
supernovae and/or AGN
\citep[e.g][]{Cattaneo:06,Croton:06,Somerville:08,Benson:10}. 
While mergers are directly observed to act on galaxies, their numbers and associated star formation seem to remain insufficient to sustain the high star 
formation rate observed at the peak in the star formation history.  
Cold gas accretion fueling star formation has therefore been proposed 
\citep{Keres:05,Sancisi:08,Dekel:09,Carilli:10}, 
but the extent of this process is yet to be confirmed from direct observational
evidence \citep{Bouche:13}.
The strong decrease in the star formation rate (SFR) density since  
$z\sim1$ \citep[e.g.,][]{Lilly:96,Madau:96,Tresse:07,Bouwens:11,Cucciati:12} 
calls for an active process to quench the star formation. 
While its origin is still unknown, it might be produced either by 
mass--dependant 
internal processes or be related to the environments in which galaxies reside. 

Each physical process that builds galaxies during cosmic time is expected to leave
specific observational signatures, even if simulations cannot reliably predict them. 
Given this context, it is necessary to seek quantitative galaxy properties that
describe how stellar mass has been assembled in the different components of
galaxies.
A key signature of galaxy evolution is the strong change in the morphological
properties, with galaxies evolving from small irregular shapes at early epochs 
to the well--structured sequence of Hubble types at the present epoch. 
Because the main components of galaxies today are bulges and discs, it is crucial to
trace the onset of these components since early times.
The evolution of the luminosity density in galactic bulge and disc components 
is a powerful method to follow galaxy build--up, which may indicate when and how 
stars have been transferred into these components. 
From a subset of the Sloan Digital Sky Survey data \citet{Tasca:11} have shown
that, averaging over the galaxy population as a whole, 
$(54 \pm 2)\%$ of the local cosmic luminosity density comes from discs and 
$(32 \pm 2)\%$ from ``pure bulge'' systems. 
Of the remaining $(14 \pm 2)\%$ half comes from the light in the 
spheroidal component of spiral galaxies and the other half from light in 
bars of systems with detectable discs.

The COSMOS survey \citep{Scoville:07a} provides a unique opportunity to make 
these measurements at about half the age of the Universe 
by combining high--resolution HST/ACS imaging data \citep{Koekemoer:07} and 
accurate spectroscopic redshifts from the zCOSMOS survey \citep{Lilly:07}.
In this letter we present for the first time an estimate of the luminosity 
functions (LF) of galaxy bulges and discs , and the relative contribution
of the associated luminosity densities (LD) of these galactic components to the
global $B$--band rest--frame LD at redshift $z\sim 0.8$.
We discuss the evolution of the fraction of 
$B$--band light in bulges and discs since $z\sim 0.8$ as well as the 
implication for the general picture of galaxy formation and evolution.
Throughout this letter we adopt a concordance 
cosmology with $\Omega_M = 0.27$, $\Omega_{\Lambda} = 0.73$ and 
$H_0= 70$ Km s$^{-1}$ Mpc$^{-1}$. All magnitudes are quoted in the AB system.


\section{Observations and data analysis}

\subsection{Sample}
\label{sec:sample}

We used the final sample of $20707$ objects, the so--called 
{\it 20k}, drawn from the zCOSMOS--bright survey \citep{Lilly:07},
a magnitude--limited spectroscopic redshift survey conducted with  
the VIMOS instrument on the ESO--VLT \citep{LeFevre:03}. 
These observations cover the full 1.7 deg$^2$ of the COSMOS HST/ACS field
\citep{Scoville:07b}, and the pure magnitude selection at $I_{AB}<22.5$ yields 
redshifts in the range $0.1<z<1.2$.
 
A detailed redshift--reliability flag system was adopted 
\citep[see][]{Lilly:09}, 
and we considerd galaxies with high--confidence redshift
measurements that translates into
a spectroscopic reliability rate of $98.6\%$ and
represents approximately $83\%$ of the whole sample.
The mean target sampling rate of the {\it 20k} sample is 
about $\sim 50\%$ \citep[see][]{Knobel:12}.
The target sampling rate of zCOSMOS objects does not depend on the size, 
brightness, or redshift; a large portion of stars was excluded from the 
spectroscopic sample based on their photometry and spectral energy 
distribution (SED).

We added several well--defined criteria to the redshift reliability to
select our final sample.
Broad--line AGNs and residual stars as well as objects with photometric defects
(less than $2\%$) were removed.
We also required morphological information from the modelling
of HST/ACS images (see section~\ref{sec:analysis}). 
Thus the final sample in this study was reduced by less than $\sim 5\%$.

To compare the morphological properties of galaxies at different cosmic epochs
in the same rest--frame band, one needs to apply a k--correction to each 
morphological component. 
This is complicated because of the different SED of the bulge and disc components 
in a galaxy and the fuzzy mixing of these components. 
To alleviate this problem, we restricted our analysis to the redshift range 
$0.7<z<0.9$, where the observed HST images in the F814W band
correspond to the rest--frame $B$ band, in which the $z\sim 0.1$ study has been 
performed \citep{Tasca:11}. 
Thus we need no k--correction, which minimised any uncertainty related to the mixing of various stellar 
populations with different SEDs.
The final sample for this study consists of $3266$ galaxies. 
The collection of deep multi--band photometry in the COSMOS field 
\citep{Taniguchi:07,Capak:07,Ilbert:09,LeFloch:09,McCracken:10,McCracken:12} enables accurate 
measurements of $B$ band absolute magnitudes, which were computed following the
method described in the appendix of \citet{Ilbert:05}. 


\subsection{Morphological measurements}
\label{sec:analysis}

We derived quantitative morphological parameters for all galaxies in our sample
using the high--resolution HST/ACS F814W images with a pixel size of 0.03
arcsec. 
We measured the bulge--to--total light fractions (B/T) using
{\it{Gim2D}} \citep{Simard:02}, a two--dimensional (2D) photometric decomposition
algorithm. We fitted each galaxy image to a superposition
of an elliptical component with a S\'ersic profile 
for the bulge and a concentric elliptical component with an 
exponential profile for the disc.
The S\'ersic index $n$, defined in \citet{Sersic:68}, controls the degree of 
curvature of the profile: a larger $n$ reflects a more centrally concentrated 
profile. In fitting the bulge component we let $0<n<10$, while we fix $n=1$ for 
fitting the disc light.

We began with the list of our source positions and used 
the {\it{SExtractor}} package \citep{Bertin:96} to perform galaxy photometry in
each sub--field to estimate the local sky background level at each point and to
define the isophotal area where each object is above the detection threshold. 
When SExtractor performs galaxy photometry, it constructs a segmentation
(or mask) image in which pixels belonging to the same
object all have the same value and sky background pixels are
flagged by zeros. Our 2D image fit was carried out on all pixels
belonging to the same SExtractor--defined object. 
Like other fitting algorithms, {\it{Gim2D}} needs initial first--guess values to 
start the computation as well as an assigned specific range for each fitting 
parameter. 
We determined these quantities with SExtractor: for each galaxy the 
photometric value computed by SExtractor (i.e. magnitude, position angle, 
half--light radius, etc.) was used as initial first--guess quantity; the range 
qas instead chosen to be the same for all the galaxies and to be wide enough
around the mean value of the distribution to allow for the various galaxy types
in our sample.

The fitting algorithm then produces values and uncertainties for each
model parameter. 
When {\it{Gim2D}} fits the 2D galaxy surface brightness to compute the galaxy 
structural parameters
it considers not only the pixels assigned to the main galaxy by the mask, but 
all pixels flagged as object or background in the SExtractor segmentation image.
Important information about the galaxy can be contained in the pixels below 
the detection threshold.
The model image of each galaxy is then convolved with a point spread function
\citep[see][]{Rhodes:07} before comparison with the real data. 

Two--dimensional bulge--to--disc decomposition of moderately distant galaxies
remains a somewhat difficult exercise. 
Due to the $(1+z)^4$ surface brightness dimming of cosmological sources, the 
disc of high--redshift galaxies can be very difficult to detect.
Some galaxy structural parameters are hence better determined than others.
To alleviate this difficulty we obtained the final galaxy flux by 
integrating 
the best--fit model over all pixels, assuring that we did not lose flux in the 
masked regions.
This guarantees that the computation of the integrated bulge and disc
luminosities, and as a consequence the B/T parameter, is quite robust. 
To determine the {\it{Gim2D}} reliability in measuring B/T we ran an extensive set of simulations that uniformly covered the whole range of B/T, galaxy magnitudes, and bulge and disc sizes. The precision of our B/T measurement is $ 10 \%$. This was taken into account when computing the bulge and disc magnitudes used to derive the corresponding luminosity functons.
Systematic errors due to the assignment of isophote twists or axial ratio 
changes to spurious thin discs were also considered and corrected for in our
analysis.
Finally, sometimes the fit did not converge or converged to a wrong
model as identified from the $\chi^2$ and the examination of the residual 
image after best fit image subtraction.   
This occurred for fewer than $5\%$ of the galaxies in our sample, and the
effect on our results was taken into consideration when we computinged the luminosity functions (LFs).

%
\begin{figure*}[!ht]
   \centering
   \includegraphics[width=5.5cm]{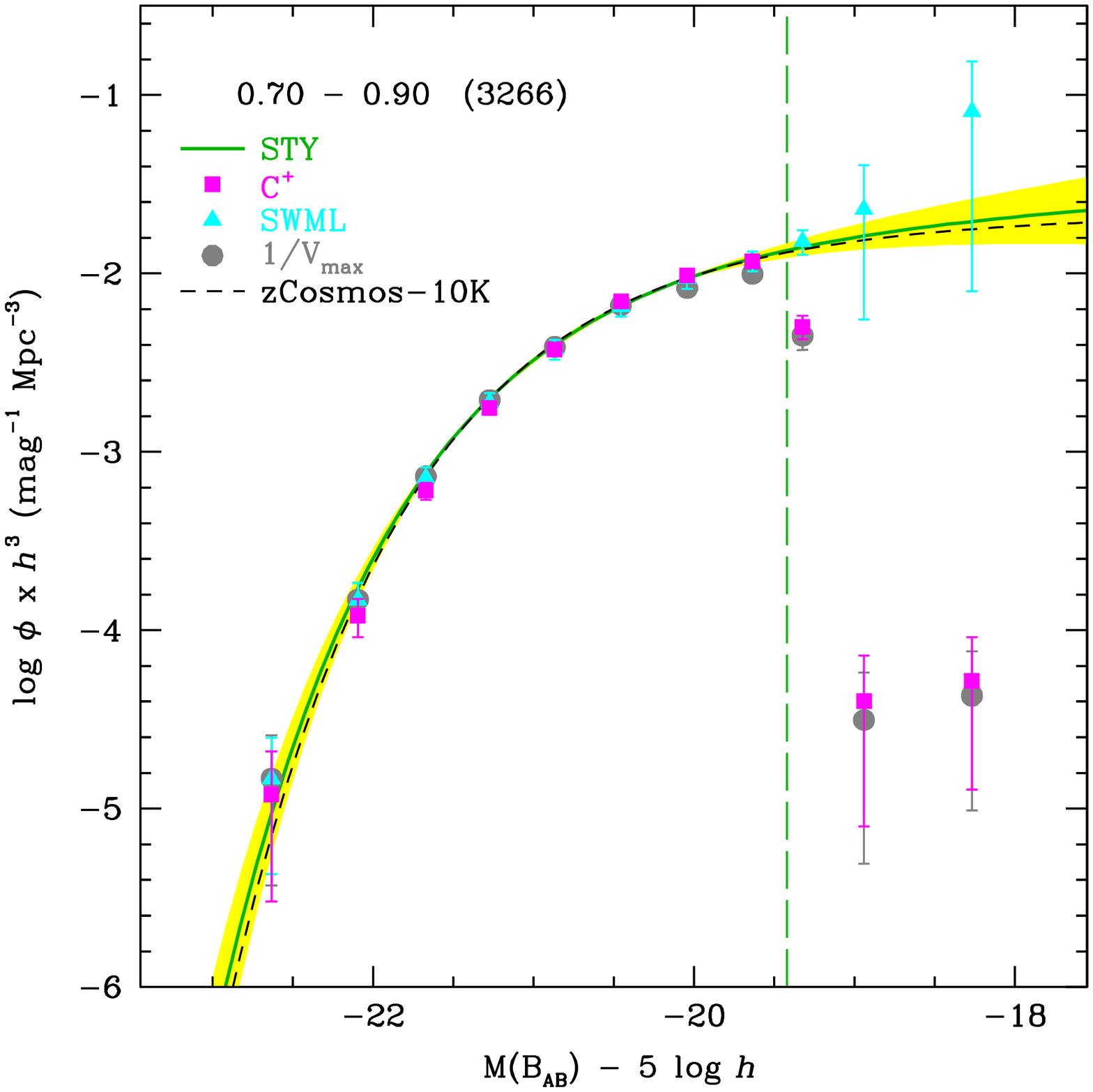}
   \includegraphics[width=5.5cm]{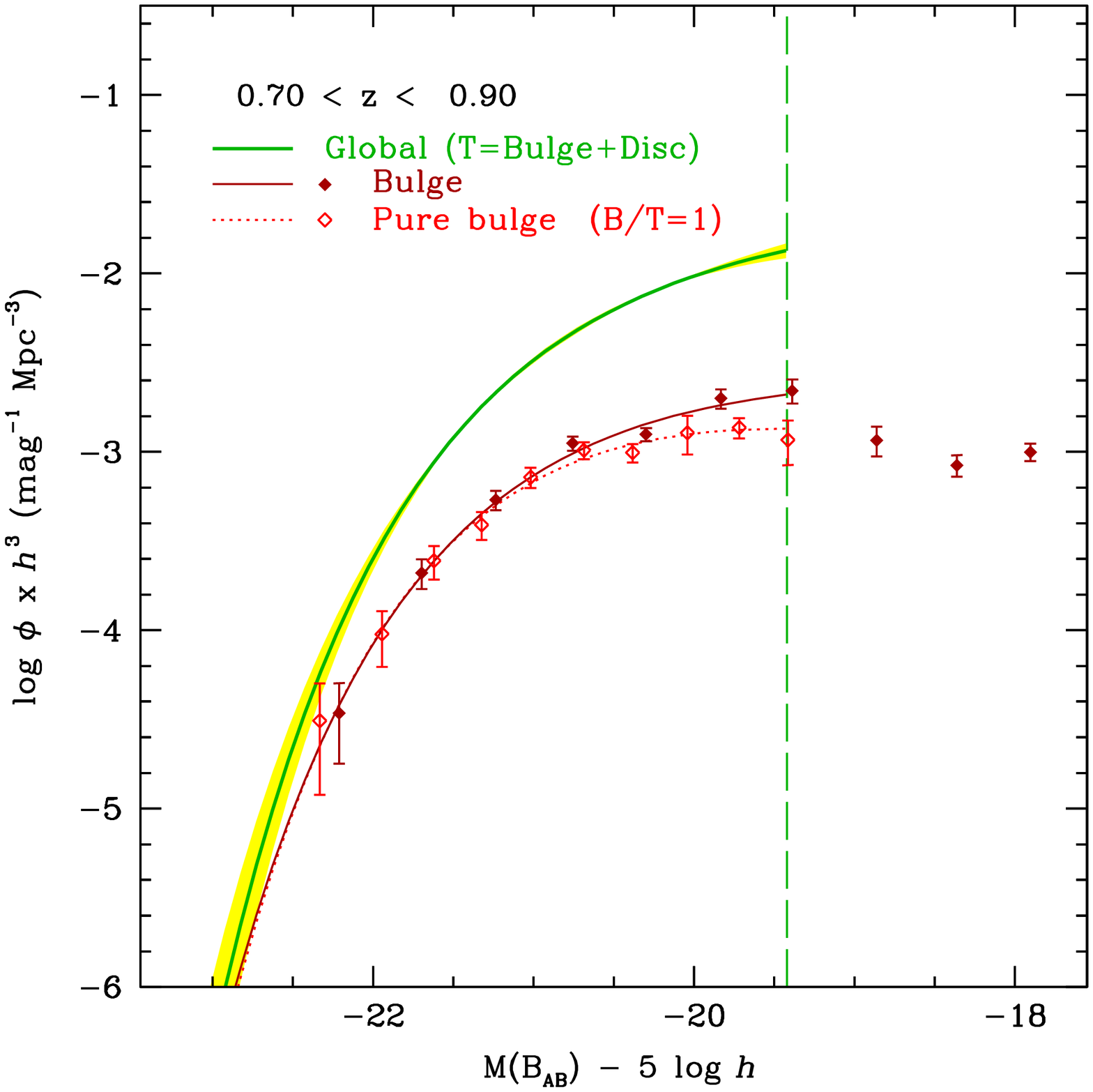}
   \includegraphics[width=5.5cm]{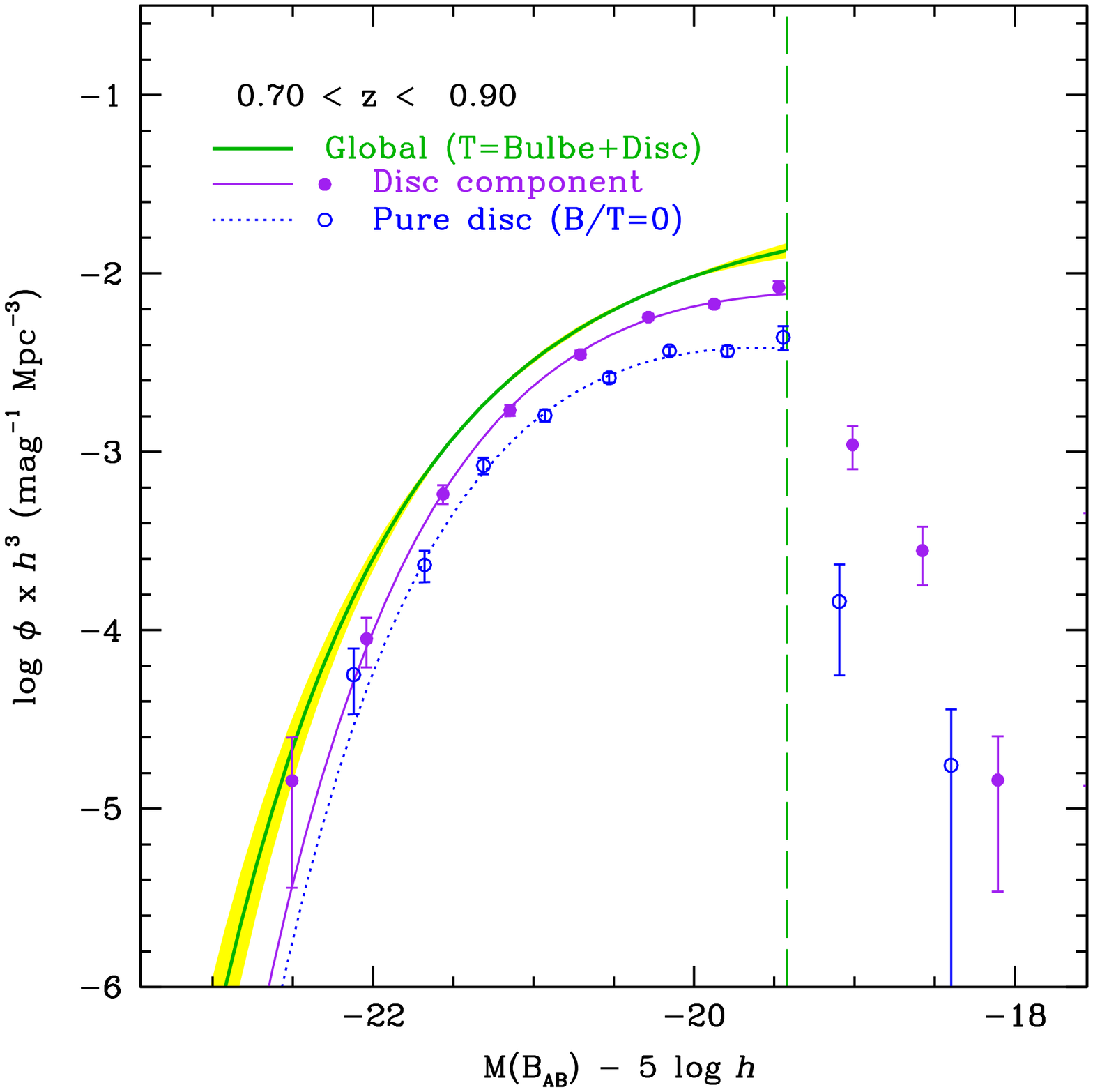}
   \caption{Left panel: Global luminosity function in the $B$--band rest--frame 
   at redshifts $0.7<z<0.9$. The total number of galaxies used is 3266. The 
   vertical dashed line represents the limit beyond which the STY estimate of
   the LF becomes incomplete. 
   The STY results for this work over the {\it 20k} sample (solid green line and
   yellow shade, also reported in the two other panels) and for the {\it 10k} sample
   (dashed black line, Zucca et al. 2009)
   are shown as well as our
   results for the $C^+$ (squares), SWML (triangles) and $1/V_{max}$ (circles).
   The shaded region represents the $68\%$ uncertainties of the parameters
   $\alpha$ and $M^{\ast}$.  
   Central panel: Galaxy bulge component (ellipticals + spiral bulges)
   and pure bulge galaxies (ellipticals) LFs at redshifts $0.7<z<0.9$ in the 
   $B$--band rest--frame.
   $1/V_{max}$ results for galaxy bulges (filled diamonds) and ellipticals
   (empty diamonds) and the respective $1/V_{max}$ fits (solid and 
   dotted lines).   
   Right panel: Galaxy disc component and and pure disc galaxies (B/T=0) LFs at
   redshifts $0.7<z<0.9$ in the $B$--band rest--frame. 
   $1/V_{max}$ results for galaxy discs (filled circles) and irregular galaxies
   (empty circles) and the respective $1/V_{max}$ fits (solid and dotted
   lines).}
   \label{fig:LF_bulge_disk}
   \end{figure*}


\section{Bulge and disc luminosity functions}
\label{sec:LF}

The LFs were computed using the algorithm for luminosity function 
\citep[{\bf{ALF}},][]{Ilbert:05}. 
We derived the LF in the redshift interval $0.7<z<0.9$, where the $I$ band
corresponds to the $B$--band rest--frame, to eliminate, or at least to strongly
minimise, any uncertainty related to differential k--corrections or to the bias
described in \citet{Ilbert:04}.

In a standard manner, to take into account unknown redshifts (for unobserved 
objects and poor--quality spectra), a weight was applied to each galaxy 
\citep[e.g.,][]{Ilbert:05}. This weight is a combination of two different 
contributions: the target sampling rate (TSR) and the spectroscopic 
success rate (SSR).
Specific to this work, we added to this weighting scheme the morphology success
rate (MSR), which is the number of spectroscopic sources with morphological 
structural parameters successfully measured, as discussed in 
section~\ref{sec:analysis}.
For each galaxy the associated final weight is given by $w_i = w^{TSR}_i \times 
w^{SSR}_i \times w^{MSR}_i$.

The total LF at redshifts $0.7<z<0.9$ is presented in the left panel of 
Figure \ref{fig:LF_bulge_disk}. Four different estimators were used and
agree down to $M_{B}=-20.2$, which we took as the limiting magnitude of this
study. 
In particular, our result (with the STY parameters: $M^{\ast}=-21.25 \pm 0.08$, 
$\phi^{\ast}=(6.53 \pm 0.6)10^{-3}Mpc^{-3}$, and $\alpha=-1.11 \pm 0.1$) is
consistent with the previous estimate on the {\it10k}
sample \citep{Zucca:09}. 
The filled points in the central and right panels of
Figure~\ref{fig:LF_bulge_disk} represent the distribution of bulge and disc
luminosities derived from the {\it{Gim2D}} fit of galaxies selected with a total luminosity down to $I_{AB}=22.5$.
The pure--disc luminosity function, represented by open circles in the right panel of Figure \ref{fig:LF_bulge_disk}, clearly decrease beyond the limiting magnitude of the the survey. 
While down to this magnitude limit both the bulge and the disc LFs are complete, below $M_{B}=-20.2$ they are lower limits to the real bulge and disc LFs, as shown by the decline in the directly computed disc LF.  
This effect arises because galaxies fainter than the 
zCOSMOS survey limit ($I_{AB} = 22.5$) are, by construction, missing. 

We additionally explored the LF of pure bulges (B/T=1), or ellipticals, 
for which all the light is in the bulge component, and of pure discs (B/T=0), 
or irregular galaxies, for which no bulge is detected.
Their LFs are shown in the central and right panels of
Figure \ref{fig:LF_bulge_disk}, where the fit of the $V_{max}$ data points are
shown for each pure population.


\section{Evolution of the $B$--band rest--frame light since $z\sim0.8$}
\label{sec:discussion}

%
   \begin{figure}
   \centering
   \includegraphics[width=8cm]{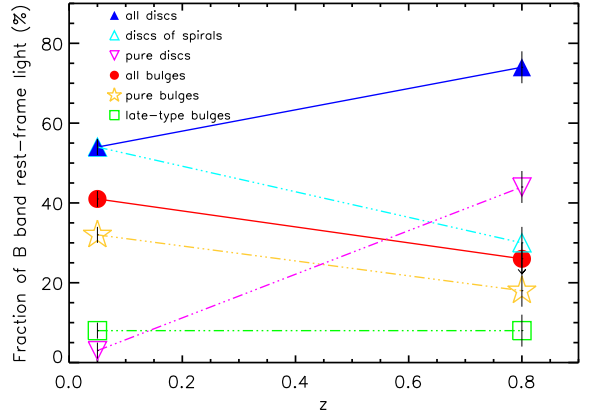}
   \caption{Evolution of the $B$--band rest--frame light since $z\sim0.8$.
   The local results at a redshiftl z~0 are taken from \citet{Tasca:11}. 
   Filled red circles and blue triangles represent the whole $B$--band 
   rest--frame light in the bulge and disc components. 
   The empty stars and squares stand for the pure bulge and the spheroid
   populations, while the empty triangles are for bulgeless
   galaxies and the disc components of spirals. 
   The lines help to guide the eyes.}
   \label{fig:ev_LD}
   \end{figure}

Using the LF presented in Section \ref{sec:LF}, 
we explored for the first time at a median redshift $\sim 0.8$ the 
fraction of $B$--band light contained in the bulge and disc components of 
galaxies brigher than $I_{AB}=22.5$. 
We computed the luminosity density (LD) as a simple sum of the $1/V_{max}$ up to
$M_{B}=-20.2$, since we are complete for this population. 
We find that for galaxies at redshifts $0.7 \leq z \leq 0.9$, $(26\pm4)\%$ of 
the $B$--band luminosity is in bulges, and $(74\pm4)\%$ in discs. 

The bulge and disc contributions of galaxies fainter than $M_{B}=-20.2$ were not investigated. We note that this represents $\sim 5\%$ of their global LD when computed as the sum of $1/V_{max}$.
Since most galaxies fainter than the bias are disc dominated and we observe that the B/T distribution evolves with magnitude towards lower values at lower luminosities, we can therefore speculate that the computed disc LD is a lower limit.

Various studies have computed the LD in bulges and discs
in the local Universe using different samples, selections, and methods
\citep[e.g.,][]{Schechter:87,Benson:07,Driver:07,Gadotti:09,Tasca:11}.
There is a general agreement that about $50\%$ of the $B$--band light in the 
local Universe is contributed by stars in discs.

In Figure\,\ref{fig:ev_LD} we compare our measurements at $z\sim0.8$ with the 
measurements of \citet{Tasca:11} at $z\sim0.1$. 
The fraction of $B$--band light in discs decreases from $(74\pm4)\%$ to
$(54\pm2)\%$, a $\sim30\%$ decrease over 6 Gyr. In contrast, 
the fraction of $B$~--band light in bulges follows a reverse evolution, 
increasing from $(26\pm4)\%$ to $(41 \pm 2)\%$ during the same period of time. 

The evolution of the fraction of light in different morphological components is 
a key signature of physical processes that shape galaxies during cosmic time, 
with the advantage of being a direct observable that does not require any 
assumption or simulation.
Our results indicate that the $B$--band emissivity has massively shifted from 
discs to bulges since $z\sim0.8$. 
Furthermore, by splitting the contribution of the bulge light into the
luminosity coming from pure bulges, meaning elliptical galaxies, and the
luminosity produced by late--type bulges, identified with the central 
component of spiral galaxies, we are able to follow their different evolution.
While elliptical galaxies and the bulges of spirals have been commonly studied
as a single population, it is now evident that the stellar populations in these two components follow a distinct evolution, which
indicates that
different physical processes must be at work.
In Figure\,\ref{fig:ev_LD} it is clearly visible that while the fraction of 
light in late--type bulges since $z\sim0.8$ remains almost constant, the
luminosity density in pure bulges increases during the same period from
$(18\pm4)\%$ to $(32\pm2)\%$, determining the global behaviour of the $B$--band
light in the bulge component.
We point out that our estimate of the fraction of light in all bulges at 
$\sim0.8$ is an upper limit because it includes the contribution of bars, 
which is out of the scope of this paper to study, while the value in the local
Universe was computed without the bar contribution, which is separately estimated to
be $(6\pm2)\%$ \citep{Tasca:11}.  
The estimate of the fraction of light in the discs of spirals and in the pure
discs, meaning bulgeless galaxies, shows that the strong evolution since 
$z\sim0.8$ of the luminosity density in the global disc component is mainly caused by 
the considerable evolution of bulgeless galaxies.

In late--type galaxies the $B$--band emissivity is tightly related to the SFR 
\citep[e.g.,][]{Tresse:02}, and therefore traces the 
on--going instantaneous star formation. 
The strong diminution of the fraction of $B$--band light in discs
is then connected to the sharp decrease of the star formation rate observed 
since $z\sim1$.  
In contrast, for bulges the $B$--band emissivity is mainly dominated by 
long--lived stars from older stellar populations instead of on--going 
starburst, and represents an integrated SFR along the time life of the bulge.
Thus the galaxy population is decreasing its SFR density within all discs, 
either via a fading of the stellar population and/or a decrease in number
density.

While at $z\sim1$ the Hubble sequence is already in place, 
\citet{Tasca:09} reported a sizeable growth of the fraction of irregular
galaxies towards higher redshifts, 
balanced by the continuous decrease of the elliptical fraction from 
$\sim30\%$ at low redshift to $\sim20\%$ at $z\sim1$. The fraction of spiral 
galaxies instead remains rather constant at $\sim50\%$.
When this morphological evolution is related to the behaviour of the evolution of the $B$--band emissivity shown in Figure\,\ref{fig:ev_LD} we conclude that while for the disc component the fading of the stellar population is not the main factor responsible for the observed trend, an important morphological change is still on--going from $z\sim1$ to $z\sim0$, mainly driven by the transformation of irregular galaxies and their strong decrease in number density and the consequent increase of the bulge component $B$--band emissivity.

We emphasise the importance of extending the analysis of bulge 
and disc components, or their progenitors, to earlier cosmic epochs to obtain 
better insight into physical processes that drive galaxy formation and 
evolution.   
Our results provide an observational reference to test theoretical model predictions.
In particular, forming bulgless galaxies have until very recently been a major challenge for hydrodynamical simulations \citep[see][]{Scannapieco:09,Marinacci:14}; our results provide the reference observed number density for these models to test against.


\begin{acknowledgements}
      This work is supported by
      funding from the European Research Council Advanced Grant ERC-2010-AdG-268107-EARLY. 
      We thank CNES and PNC for their financial support of the COSMOS project. 
      This work benefited from support from the French National Computing Centre
      (CINES), for providing part of the necessary computational resources.
      The zCOSMOS \& HST-COSMOS survey data used in this work 
      have been obtained from the databases operated by Cesam, 
      Laboratoire d'Astrophysique de Marseille, France. 
 
\end{acknowledgements}


\bibliographystyle{aa} 
\bibliography{paperbib} 


\end{document}